\documentclass{cernphprep}
\usepackage{xcolor,pict2e,curve2e}

\newcommand{\Eq}[1]   {Eq.~(\ref{#1})}
\newcommand{\Eqs}[2]  {Eqs.~(\ref{#1}) and~(\ref{#2})}
\newcommand{\Eqr}[2]  {Eqs.~(\ref{#1})--(\ref{#2})}
\newcommand{\Fi}[1]   {Fig.~\ref{#1}}

\newcommand{\gevc}    {\mbox{GeV$/c$}}
\newcommand{\gevcc}   {\mbox{GeV$/c^2$}}

\newcommand{\mevcc}   {\mbox{MeV$/c^2$}}
\newcommand{\rb}[1]   {\mbox{\textrm{\scriptsize #1}}}
\newcommand{\rbt}[1]  {\mbox{\textrm{\tiny #1}}}
\newcommand{\vzero}   {\ensuremath{\textrm{V}^{0}}}
\newcommand{\lam}     {\ensuremath{\Lambda}}
  
\newcommand{\xim}     {\ensuremath{\Xi^{-}}}
\newcommand{\xizero}  {\ensuremath{\Xi^{0}}}

\newcommand{\sig}     {\ensuremath{\Sigma^{0}}}
\newcommand{\sip}     {\ensuremath{\Sigma^{+}}}

\newcommand{\pimin}   {\ensuremath{\pi^-}}

\newcommand{\kmin}    {\ensuremath{\textrm{K}^-}}
\newcommand{\kplus}   {\ensuremath{\textrm{K}^+}}

\newcommand{\lamp}    {\ensuremath{\textrm{p}-\Lambda}}

\newcommand{\sqrts}   {\ensuremath{\sqrt{s_{_{\rbt{NN}}}}}}
\newcommand{\pt}      {\ensuremath{p_{\rb{t}}}}
\newcommand{\mt}      {\ensuremath{m_{\rb{t}}}}
\newcommand{\kt}      {\ensuremath{k_{\rb{t}}}}

\newcommand{\mzero}   {\ensuremath{m_{\rb{0}}}}
\newcommand{\mtavg}   {\ensuremath{\langle m_{\rb{t}} \rangle}}

\newcommand{\qinv}    {\ensuremath{q_{\rb{inv}}}}
\newcommand{\dedx}    {\ensuremath{\textrm{d}E/\textrm{d}x}}

\newcommand{\nwound}  {\ensuremath{\langle N_{\rb{w}} \rangle}}

\newcommand{\betat}   {\ensuremath{\beta_{\rb{t}}}}
\newcommand{\etaf}    {\ensuremath{\eta_{\rb{f}}}}
\newcommand{\tauzero} {\ensuremath{\tau_{\rb{0}}}}

\newcommand{\dzeros}  {\ensuremath{d_{\rb{0}}^{\rb{S}}}}
\newcommand{\dzerot}  {\ensuremath{d_{\rb{0}}^{\rb{T}}}}
\newcommand{\fzeros}  {\ensuremath{f_{\rb{0}}^{\rb{S}}}}
\newcommand{\fzerot}  {\ensuremath{f_{\rb{0}}^{\rb{T}}}}
\newcommand{\chisq}   {\ensuremath{\chi^{2}}}

\newcommand{\rgaus}   {\ensuremath{R_{\rb{G}}}}
\newcommand{\rgeo}    {\ensuremath{R_{\rb{geo}}}}
\newcommand{\rout}    {\ensuremath{R_{\rb{out}}}}
\newcommand{\rside}   {\ensuremath{R_{\rb{side}}}}
\newcommand{\rlong}   {\ensuremath{R_{\rb{long}}}}

\begin{document}
\begin{titlepage}
\PHnumber{2011-032}
\PHdate{16 March 2011}
\title{Proton -- Lambda Correlations in Central Pb+Pb Collisions at
  \sqrts~=~17.3~GeV}
\begin{Authlist}
T.~Anticic\Iref{22}, 
B.~Baatar\Iref{8},
D.~Barna\Iref{4},
J.~Bartke\Iref{6},
H.~Beck\Iref{9},
L.~Betev\Iref{10},
H.~Bia{\l}\-kowska\Iref{19}, 
C.~Blume\Iref{9},
M.~Bogusz\Iref{21},
B.~Boimska\Iref{19},
J.~Book\Iref{9},
M.~Botje\Iref{1},
P.~Bun\v{c}i\'{c}\Iref{10},
T.~Cetner\Iref{21},
P.~Christakoglou\Iref{1},
P.~Chung\Iref{18},
O.~Chv\'{a}la\Iref{14},
J.G.~Cramer\Iref{15},
V.~Eckardt\Iref{13},
Z.~Fodor\Iref{4},
P.~Foka\Iref{7},
V.~Friese\Iref{7},
M.~Ga\'{z}dzicki\Iref{9,11},
K.~Grebieszkow\Iref{21},
C.~H\"{o}hne\Iref{7},
K.~Kadija\Iref{22},
A.~Karev\Iref{10},
V.I.~Kolesnikov\Iref{8},
M.~Kowalski\Iref{6},
D.~Kresan\Iref{7},
A.~L\'{a}szl\'{o}\Iref{4},
R.~Lacey\Iref{18},
M.~van~Leeuwen\Iref{1},
M.~Ma\'{c}kowiak\Iref{21},
M.~Makariev\Iref{17},
A.I.~Malakhov\Iref{8},
M.~Mateev\Iref{16},
G.L.~Melkumov\Iref{8},
M.~Mitrovski\Iref{9},
St.~Mr\'{o}wczy\'{n}ski\Iref{11},
V.~Nicolic\Iref{22},
G.~P\'{a}lla\Iref{4},
A.D.~Panagiotou\Iref{2},
W.~Peryt\Iref{21},
J.~Pluta\Iref{21},
D.~Prindle\Iref{15},
F.~P\"{u}hlhofer\Iref{12},
R.~Renfordt\Iref{9},
C.~Roland\Iref{5},
G.~Roland\Iref{5},
M.~Rybczy\'{n}ski\Iref{11},
A.~Rybicki\Iref{6},
A.~Sandoval\Iref{7},
N.~Schmitz\Iref{13},
T.~Schuster\Iref{9},
P.~Seyboth\Iref{13},
F.~Sikl\'{e}r\Iref{4},
E.~Skrzypczak\Iref{20},
M.~S{\l}odkowski\Iref{21},
G.~Stefanek\Iref{11},
R.~Stock\Iref{9},
H.~Str\"{o}bele\Iref{9},
T.~Susa\Iref{22},
M.~Szuba\Iref{21},
M.~Utvi\'{c}\Iref{9},
D.~Varga\Iref{3},
M.~Vassiliou\Iref{2},
G.I.~Veres\Iref{4},
G.~Vesztergombi\Iref{4},
D.~Vrani\'{c}\Iref{7},
Z.~W{\l}odarczyk\Iref{11},
A.~Wojtaszek-Szwarc\Iref{11},
\end{Authlist}
\Instfoot{1}{NIKHEF, 
             Amsterdam, Netherlands.}
\Instfoot{2}{Department of Physics, University of Athens, 
             Athens, Greece.}
\Instfoot{3}{E\"{o}tv\"{o}s Lor\'{a}nt University, 
             Budapest, Hungary.}
\Instfoot{4}{KFKI Research Institute for Particle and Nuclear Physics,
             Budapest, Hungary.} 
\Instfoot{5}{MIT, 
             Cambridge, USA.}
\Instfoot{6}{H. Niewodnicza\'{n}ski Institute of Nuclear Physics, 
             Polish Academy of Sciences, 
             Cracow, Poland.}
\Instfoot{7}{Gesellschaft f\"{u}r Schwerionenforschung (GSI),
             Darmstadt, Germany.} 
\Instfoot{8}{Joint Institute for Nuclear Research, 
             Dubna, Russia.}
\Instfoot{9}{Fachbereich Physik der Universit\"{a}t, 
             Frankfurt, Germany.}
\Instfoot{10}{CERN, 
             Geneva, Switzerland.}
\Instfoot{11}{Institute of Physics, Jan Kochanowski University, 
             Kielce, Poland.}
\Instfoot{12}{Fachbereich Physik der Universit\"{a}t, 
             Marburg, Germany.}
\Instfoot{13}{Max-Planck-Institut f\"{u}r Physik, 
             Munich, Germany.}
\Instfoot{14}{Inst. of Particle and Nuclear Physics, Charles Univ.,
             Prague, Czech Republic.}
\Instfoot{15}{Nuclear Physics Laboratory, University of Washington,
             Seattle, WA, USA.} 
\Instfoot{16}{Atomic Physics Department, Sofia Univ. St.~Kliment Ohridski, 
             Sofia, Bulgaria.}
\Instfoot{17}{Institute for Nuclear Research and Nuclear Energy, BAS, 
             Sofia, Bulgaria.}
\Instfoot{18}{Department of Chemistry, Stony Brook Univ. (SUNYSB), 
             Stony Brook, USA.}
\Instfoot{19}{Institute for Nuclear Studies, 
             Warsaw, Poland.}
\Instfoot{20}{Institute for Experimental Physics, University of Warsaw,
             Warsaw, Poland.} 
\Instfoot{21}{Faculty of Physics, Warsaw University of Technology, 
             Warsaw, Poland.}
\Instfoot{22}{Rudjer Boskovic Institute, 
             Zagreb, Croatia.}
\ShortAuthor{T.~Anticic \emph{et al.}}
\Collaboration{NA49 collaboration}
\vfill

\begin{abstract}
The momentum correlation between protons and lambda particles emitted
from central Pb+Pb collisions at \sqrts~= 17.3~GeV was studied by the
NA49 experiment at the CERN SPS.  A clear enhancement is observed for
small relative momenta ($\qinv < 0.2$~GeV).  By fitting a theoretical
model, which uses the strong interaction between the proton and the
\lam\ in a given pair, to the measured data a value for the effective
source size is deduced.  Assuming a static Gaussian source
distribution we derive an effective radius parameter of $\rgaus = 3.02
\pm 0.20$(stat.)$^{+0.44}_{-0.16}$(syst.)~fm. 
\end{abstract}

\vspace{2cm}
\Submitted{(Submitted to Phys. Rev. C)}

\end{titlepage}

\section{Introduction}

Heavy-ion collisions at ultra-relativistic energies produce strongly 
interacting matter under extreme conditions.  The main goal is to
create in these reactions a state in which the confinement of quarks
and gluons inside hadrons is no longer effective, the so-called
quark-gluon plasma.  This strongly compressed matter undergoes a rapid
expansion with a drop of temperature and energy density.  Two-particle
momentum correlations provide unique information on the size and
dynamic evolution of this fireball and are therefore a widely employed
observable in heavy-ion physics.  Usually correlations of identical
charged pions are studied which, due to the high available statistics,
allow a multi-dimensional study of radius parameters \cite{PRATT86,
BERTSCH89,NA49HBT}.  Less frequently two-proton correlations are
analyzed \cite{NA49PP,NA44PP,WA80PP,E895PLAM}, although only
one-dimensionally.  Moreover, the large abundance of strange particles
produced in heavy-ion collisions allows to study also two particle
correlations between pairs of strange particles or pairs of strange
and non-strange particles.  For example the correlations of identical
kaons were investigated at the CERN-SPS \cite{NA49KK} as well as at
RHIC \cite{STARKK}.  In this paper we report on the measurement of the
\lamp\ correlation function in central Pb+Pb collisions at \sqrts~=
17.3~GeV at the CERN-SPS.  

It was suggested that also the momentum correlation between \lam\ and
protons can be employed to measure the size of the emitting source
\cite{WANG99}.  The correlation function of \lamp\ pairs is only
affected by the strong interaction between the particles.  This
distinguishes \lamp\ correlations from the two-proton case, for which
the correlation function is dominated by the repulsive Coulomb
interaction and the Fermi-Dirac statistics at low relative
momenta. Both effects are absent in the \lamp\ correlation function,
which should therefore be more sensitive to large source
sizes \cite{WANG99}.  However, the knowledge of the strong
interaction between protons and \lam\ is necessary to relate the
strength of the correlation to the size of the emitting source.  There
is a substantial set of data available on low-energy elastic $\lam$~p
scattering \cite{ALEXANDER68,SECHIZORN68,KADYK71}, and on
\mbox{$\kmin \textrm{d} \rightarrow \lam \: \textrm{p} \: \pimin$}
\cite{DAHL61,CLINE68,BRAUN77}, as well as 
\mbox{$\textrm{p} \textrm{p} \rightarrow \textrm{p} \kplus \lam$}
\cite{COSY11,HIRES} reactions.  Also \lam\ hypernuclei provide
important information on the \lam\ nucleon interaction.  Based on this
data many theoretical analyses derived \lamp\ scattering lengths and
effective interaction ranges \cite{NAGELS79,BODMER88,MAESSEN89,
RIJKEN99,HAIDENBAUER05,POLINDER06,SIBIRTSEV06} which can be used to
calculate the \lamp\ correlation function.

A preliminary study by the NA49 experiment at the SPS was
reported in \cite{BLUME02}.  Here we describe the final results of a 
new and improved analysis \cite{BECK09}.  Similar studies of \lamp\
correlations in heavy ion collisions were performed at lower
\cite{HADESPLAM,E895PLAM} and higher \cite{STARPLAM} center-of-mass
energies, allowing to investigate the evolution of \lamp\ correlations
with \sqrts.


\section{Data Analysis}

The data presented here were measured by the NA49 experiment at 
the CERN SPS. A detailed description of the experimental setup can be 
found in \cite{NA49NM}.  Charged particles produced by interactions
of the Pb beam in a thin Pb-foil target are tracked with four 
large-volume Time Projection Chambers (TPCs).
Two TPCs are placed inside two superconducting dipole
magnets, while the other two are situated outside of the magnetic
field.  Since the latter measure long pieces of the particle tracks,
they allow a precise determination of the specific energy loss \dedx\
inside the detector gas (typical resolution of 4~\%) and thus particle
identification in a large region of phase-space.  Additional particle
identification is provided around mid-rapidity by Time-Of-Flight
detectors.  A Zero Degree Calorimeter is used to measure the energy in
the projectile fragmentation region from which the centrality of the
reaction can be deduced.  This analysis is based on \mbox{$2.8 \cdot
  10^{6}$} Pb+Pb events at \sqrts~= 17.3~GeV recorded in the year
2000, which cover the 23.5~\% most central part of the total inelastic
cross section,  corresponding to an averaged number of wounded
nucleons of $\nwound = 262$.

\subsection{\lam\ reconstruction}

The \lam\ hyperons are detected via their charged decay 
\mbox{$\Lambda \rightarrow \textrm{p} \pimin$}, using the same methods
as described in \cite{NA49EHYP,NA49SHYP}.  The \lam\ reconstruction is
done by forming pairs of positively and negatively charged tracks and
extrapolating them towards the main interaction vertex.  The
positively (negatively) charged tracks are required to have at least
50 (30) reconstructed points.  Pairs with a distance of closest
approach of less than 0.5~cm anywhere between the position of the
first measured point on the tracks and the target plane are considered
as \vzero\ candidates.  Assigning proton and pion masses to the
positively and negatively charged decay particle, the invariant mass
of a \lam\ candidate is calculated.  A significant reduction of the
combinatorial background can be achieved by applying several selection
criteria to the \lam\ candidates.  In this analysis it is required
that the secondary vertex is separated by at least 25~cm in
beam-($z$)-direction from the target plane.  Additionally, the 
back-extrapolation of the flight path of the \lam\ candidate must not
deviate from the interaction vertex position in the transverse
directions $x$ and $y$ by more than $|\Delta x| = 0.75$~cm and
$|\Delta y| = 0.375$~cm.  The signal-to-background ratio is further
improved by enriching the protons in the sample of positively charged
tracks by applying a momentum-dependent cut on the measured energy
loss (\dedx).  An additional \dedx\ cut on the negatively-charged
tracks also allows to reject electrons from photon conversions.

%
\begin{figure}[t]
\includegraphics[height=80mm]{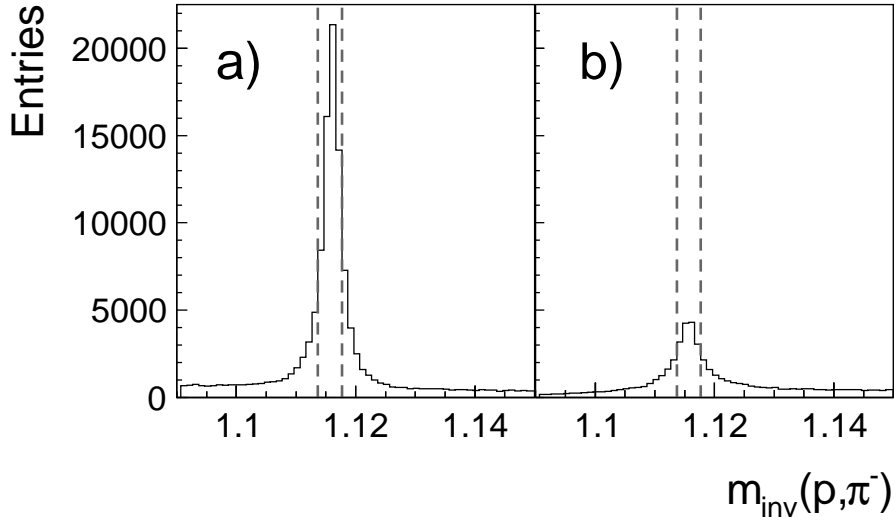}
\caption{\label{fig:lam_signal} 
The invariant mass distribution of \lam\ candidates in two exemplary
phase-space bins ((a) $-1.2 < y < -0.4$ and $0.6 < \pt < 1.2$~\gevc,
(b) $-0.4 < y < 0.4$ and $1.8 < \pt < 2.4$~\gevc) for central Pb+Pb
reactions at \sqrts~= 17.3~GeV.  The two vertical lines indicate the
mass windows used to define \lam\ candidates.}
\end{figure}
%

An important point with respect to correlation studies is the
requirement that each \lam\ candidate must be unique.  If it happens
that a daughter track of a given \lam\ candidate is also assigned to
another one, a strong artificial correlation between both candidates
is created which in turn affects the measured \lamp\ correlation
function.  In this analysis it is therefore ensured that any given
track is used only once as a daughter track.  Similarly, \lam\
daughters that were not reconstructed as a single track, but as two
track pieces (split tracks), will cause a distortion of the measured
correlation function.  To exclude these tracks it is required that the
number of measured points of each accepted track is higher than 50~\%
of the number of points that this track could maximally have according
to its trajectory in the TPCs.

Figure~\ref{fig:lam_signal} shows the distribution of invariant mass
$m_{inv}$ of \lam\ candidates obtained after assigning the proton
($\pi^{-}$) mass to  the positive (negative) daughter track for two
intervals of the center-of-mass rapidity $y$ and transverse momentum
$p_t$.  \lam\ are accepted in a mass window \mbox{$[\mzero - \Delta
  m, \mzero + \Delta m]$} of a half-width of \mbox{$\Delta m$ =
  2~\mevcc}, where \mbox{\mzero\ = 1.115683~\gevcc} is the literature
value for the \lam\ mass \cite{PDG08}.  This mass window was chosen in
order to optimize the signal-to-background ratio and reduce the
corrections for the signal purity.

%
\begin{figure}[t]
\includegraphics[height=80mm]{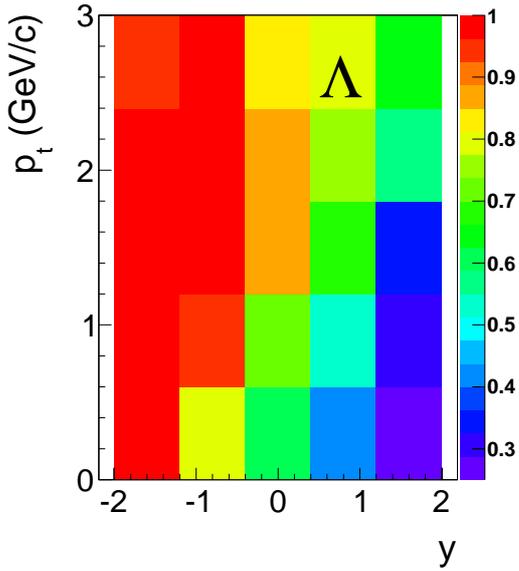}
\caption{\label{fig:purity} 
The purity $P_{\Lambda}(y,\pt)$ of the selected \lam\ candidates as
function of rapidity $y$ and transverse momentum \pt.
}
\end{figure}
%

The ratio between signal and background inside this mass window and
with it the purity of the \lam\ sample vary over phase-space.  A
$\chi^{2}$-fit to the invariant mass distributions in different $y$
and \pt\ bins was performed in order to determine the relative
contributions.  The fit function used is the sum of a polynomial for the
combinatorial background and a function for the \lam\ signal.  The
measured shape of the \lam\ signal results from a convolution of the
resolutions for the single track momenta and the secondary vertex
positions and is found to be well described by an asymmetric
Lorentz-curve.  By subtracting the background function from the
measured invariant mass distribution the number of real \lam\ is
determined.  Defining the purity $P_{\Lambda}$ by the ratio of the
number of real \lam\ to the number of accepted \lam\ candidates in the
chosen mass window one obtains the result shown in \Fi{fig:purity}.
The phase-space averaged value of the \lam\ purity is $\langle
P_{\Lambda} \rangle = 69$~\%.

%
\begin{figure}[t]
\includegraphics[height=80mm]{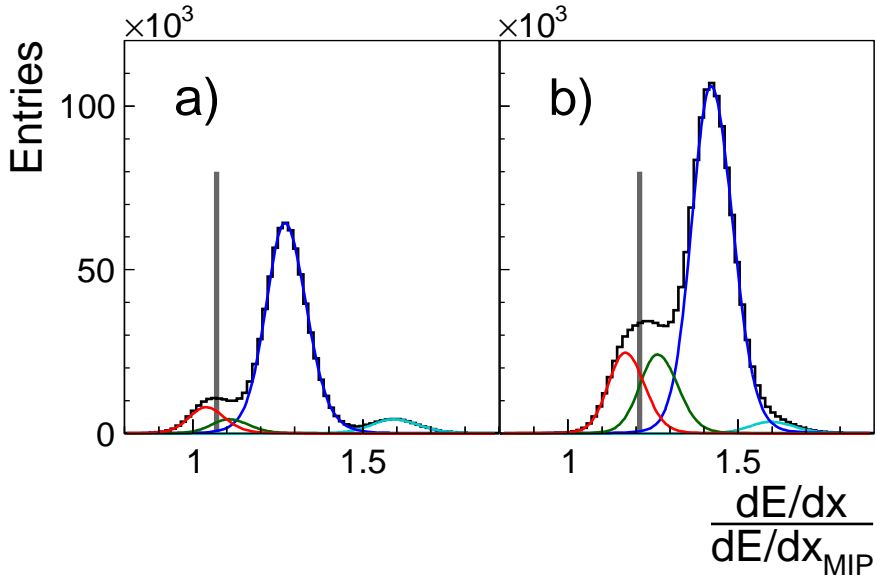}
\caption{\label{fig:dedx_fits} 
The \dedx\ spectra in two exemplary total momentum bins ((a) $4.0 < p
< 5.0$~\gevc, (b) $12.6 < p < 15.9$~\gevc) together with their
decomposition into contributions from protons, kaons, pions, and
electrons (from left to right).  Shown is the measured energy loss
normalized to the minimum ionizing value $\dedx_{\rb{MIP}}$.  The
vertical lines represent the upper \dedx\ cuts that result in a purity
of $P_{\rb{p}}= 0.8$. 
}
\end{figure}
%

\subsection{Proton reconstruction}

Protons are identified via their energy loss \dedx\ as measured in
the two Main-TPCs.  A valid proton track is required to have at least
50 reconstructed points.  An additional cut on the impact parameters
in the target plane ($|b_{\rb{x}}| < 5.0$~cm, $|b_{\rb{y}}| < 2.0$~cm)
reduces the contribution from secondary tracks.  In order to assign the 
probability of being a proton to a given track, the energy loss
spectra measured in bins of total momentum $p$ are fitted by a sum
$E(x,p)$ of asymmetric Gaussians using a \chisq\ minimizing
procedure~\cite{MILICA}:
\begin{equation}
E(x,p) = \sum\limits_{i=\rb{d,p,K,}\pi,\rb{e}} A_{i}(p) \:
         \frac{1}{\sum_{l} n_{l}} \:
         \sum\limits_{l} \frac{n_{l}}{\sqrt{2 \pi} \: \sigma_{i,l}(p)}
         \exp \left[ - \frac{1}{2} \left( 
            \frac{x - \hat{x}_{i}(p)} {(1 \pm \delta)\:\sigma_{i,l}(p)} 
            \right)^{2} \right] .
\end{equation}
Here, $A_{i}(p)$ denotes the yield for particle type $i$, $n_{l}$ the 
number of tracks in a given track length interval $l$,
$\hat{x}_{i}(p)$ the most probable \dedx\ values for particle type
$i$, $x$ the measured \dedx\ value of the track under consideration,
$\sigma_{i,l}(p)$ the width of the Gaussian, and $\delta$ the
asymmetry parameter.  The parameters $A_{i}(p)$, $\hat{x}_{i}(p)$, and
$\sigma_{\pi}(p)$ are determined by the fitting procedure in each
$p$~bin separately.  The widths for the other particles types and
different track length bins are derived from the parameter
$\sigma_{\pi}(p)$ for pions: $\sigma_{i,l}(p) = \sigma_{\pi}(p) \cdot
(\hat{x}_{i}/\hat{x}_{\pi})^{\alpha}(1/\sqrt{l})$.  The exponent was
determined to be $\alpha = 0.625$ and the parameter $\delta$ is
fixed to 0.065.  Results of the fit procedure for two different total
momentum bins are shown in \Fi{fig:dedx_fits}.  Since for low particle
momenta the energy loss curves of different particle species cross
each other and particle identification is thus not possible, a lower
cut on the total momentum of $p > 4$~\gevc\ is applied.  To exclude
the region of the Fermi plateau also momenta above 50~\gevc\ are
discarded.  Based on these fits a momentum dependent cut on the
measured \dedx\ values for single tracks is defined such that the
accepted tracks always have the same probability of being a proton.
For the standard analysis this probability is set to 80~\%, equivalent
to a constant proton purity $P_{\rb{p}}= 0.8$.  As in the case of the
\lam\ decay daughters it is ensured that split tracks are removed from
the track sample by rejecting tracks that have less than 55~\% of the
number of geometrically possible points. 

\subsection{Determination of the \lamp\ correlation function}

The selected \lam\ and proton candidates are then combined to form
\lamp\ pairs.  In order to avoid trivial auto-correlations a track
that is used to reconstruct the \lam\ candidate is removed from the
primary proton sample.  The pair distribution $S(\qinv)$ is measured
as a function of the generalized invariant relative momentum of the
\lamp\ pair \qinv, which is defined as the modulus of \mbox{$\tilde{q}
  = q - P (q P) / P^{2}$}, with $q = p_{\rb{p}} - p_{\lam}$, $P =
p_{\rb{p}} + p_{\lam}$, and $q P = m_{\rb{p}}^{2} - m_{\lam}^{2}$.
Here, $p_{\rb{p}}$ and $p_{\lam}$ are the 4-momenta of the proton and
the \lam.  In the two particle center-of-mass system $\tilde{q}$
reduces to $\{0, 2 \vec{k}^{*}\}$, with $2 \vec{k}^{*}$ being the
3-momentum difference in this reference frame \cite{LEDNICKY05}.
 
An event-mixing method that combines proton and \lam\ candidates taken
from different events, is employed for the construction of the
uncorrelated background $B(\qinv)$.  The measured correlation function
is thus defined as:
\begin{equation}
\label{eq:meascf}
C_{\rb{meas}}(\qinv) = N \: \frac{S(\qinv)}{B(\qinv)} .
\end{equation}
The normalization constant $N$ is determined by requiring
$C_{\rb{meas}}(\qinv) = 1$ in the region $0.2 < \qinv < 0.3$~GeV.
Since the reconstruction of real pairs is affected by the limited
two-track resolution of the detector \cite{NA49NM}, a distance cut
between the track of the primary proton and the track of the positive
\lam\ decay particle is applied for real pairs as well as for the
mixed event pairs.  For each pair it is required that the tracks have
an average separation of at least 3.0~cm.  This average is determined
as the arithmetic mean of the track distances determined in planes
perpendicular to the beam axis.  For each TPC two planes are taken
into account.  Their distances to the target plane are 160.5~cm,
240.5~cm, 540.0~cm, 620.0~cm, 910.0~cm, and 990.0~cm.  Pairs that do
not pass the distance cut are discarded.

%
\begin{figure}[t]
\includegraphics[height=80mm]{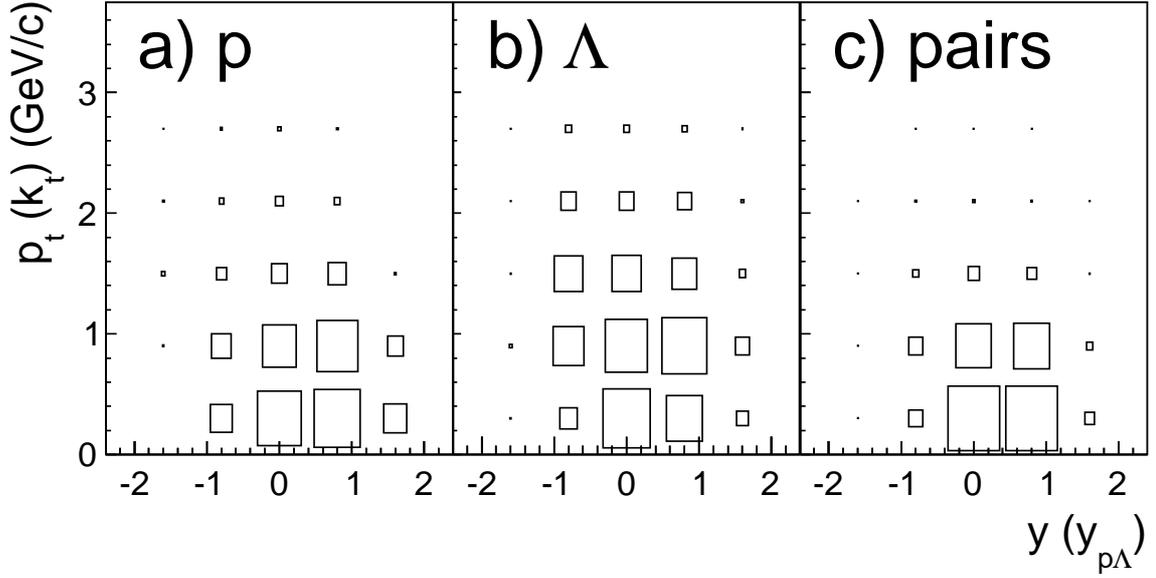}
\caption{\label{fig:acceptance} 
The phase-space population of protons (a), $\Lambda$ (b) as a function
of rapidity $y$ and transverse momentum \pt.  The distribution for the
resulting \lamp\ pairs (c) is shown versus pair rapidity 
$y_{\rb{p}\lam} = \frac{1}{2} \ln \frac{E_{\rb{p}} + E_{\lam} +
p_{\rb{z,p}} + p_{\rb{z},\lam}}{E_{\rb{p}} + E_{\lam} - p_{\rb{z,p}} -
p_{\rb{z},\lam}}$ and $\kt = \frac{1}{2} | \vec{p}_{\rb{t,p}} +
\vec{p}_{\rb{t},\lam} |$.  
}
\end{figure}
%

Figure~\ref{fig:acceptance} shows the phase-space population of the
accepted protons, \lam, and \lamp\ pairs.  Averaging over all measured
\lamp\ pairs we find \mbox{$\langle \kt \rangle = 0.53$~\gevc} with
$\kt = \frac{1}{2} | \vec{p}_{\rb{t,p}} + \vec{p}_{\rb{t},\lam} |$ 
(\mbox{$\langle \mt \rangle = 1.18$~GeV} with $\mt = \sqrt{\kt^{2} +
 (\frac{1}{2}(m_{\lam} + m_{\rb{p}}))^{2}}$).

The measured correlation function can be affected by the finite
momentum resolution of the detector.  In \cite{NA49HBT} an extensive
investigation of its influence on the radius parameters extracted from
correlations of identical charged pions as measured with the NA49
experiment is discussed.  Due to the excellent momentum resolution of
the NA49 detector, it turned out that the impact is negligible and a
correction for this effect is not necessary.  Even though the momentum
resolution for a \lam\ ($\langle \sigma_{p_{\lam}} \rangle \leq 1 \%$)
is worse than for the primary track, the resulting effect
on the measured \lamp\ correlation function is still clearly smaller
than all other systematics effects.  Therefore no correction is
applied in this analysis.

A substantial fraction of the measured protons and \lam\ originate
from weak and electro-magnetic decays of heavier particles
(feed-down).  In the following it will be assumed that these decay
particles are not correlated, because the decays happen long after the
thermal freeze-out, and will thus reduce the observed \lamp\
correlation function.  For the \lam\ the feed-down originates from
\xim, \xizero, and \sig\ decays, while in case of the protons the
decays of \lam, and \sip\ contribute.  We calculate the fraction of
protons and \lam\ originating from feed-down ($F_{\rb{p}}(y,\pt)$ and
$F_{\Lambda}(y,\pt)$) via a simulation procedure.  \lam, \xim, and
\xizero\ are generated according to their measured phase-space
distributions \cite{NA49EHYP}.  The measured \lam\ include the \sig\
which cannot be separated experimentally.  For the \xizero\ the same
input distributions are assumed as for the \xim\, scaled by the ratio
\xizero/\xim\ of the total multiplicities taken from statistical model
fits \cite{BECATTINI06}.  The daughter tracks of the generated
particles are followed through the NA49 detector setup using the
Geant3.21 package \cite{GEANT}.  The response of the TPCs to the
traversing particles is simulated with NA49 specific software.  In a next
step the simulated raw signals are added to measured raw data and
processed by the same reconstruction program as used for the
experimental data.  By applying the same cuts as in the normal
analysis, the feed-down contributions to the measured \lam\ and
protons are determined.  The $y$ and \pt\ dependences of the
feed-down fractions $F_{\rb{p}}(y,\pt)$ and $F_{\Lambda}(y,\pt)$ are
summarized in \Fi{fig:feeddown}.  The phase-space averaged values are
$\langle F_{\rb{p}} \rangle = 22$~\% and $\langle F_{\Lambda} \rangle
= 43$~\%.

The final \lamp\ correlation function $C_{\rb{corr}}(\qinv)$ results
from the measured $C_{\rb{meas}}(\qinv)$ by applying a combined
correction factor $\langle K(\qinv) \rangle$ for purity and feed-down.
This factor is determined by averaging the product
$P_{\rb{p}}(y_{\rb{p}},p_{\rb{t,p}}) \, P_{\lam}(y_{\lam}, p_{\rb{t},\lam}) \:
(1 - F_{\rb{p}}(y_{\rb{p}},p_{\rb{t,p}})) \, (1 - F_{\lam}(y_{\lam},
p_{\rb{t},\lam}))$ over all reconstructed \lamp\ pair combinations
falling into a given bin of relative momentum \qinv.  The corrected
correlation function thus follows from:
\begin{equation}
\label{eq:corrcf}
C_{\rb{corr}}(\qinv) = \frac{C_{\rb{meas}}(\qinv) - 1}
                         {\langle K(\qinv) \rangle} \: + 1 .
\end{equation}

%
\begin{figure}[t]
\includegraphics[height=80mm]{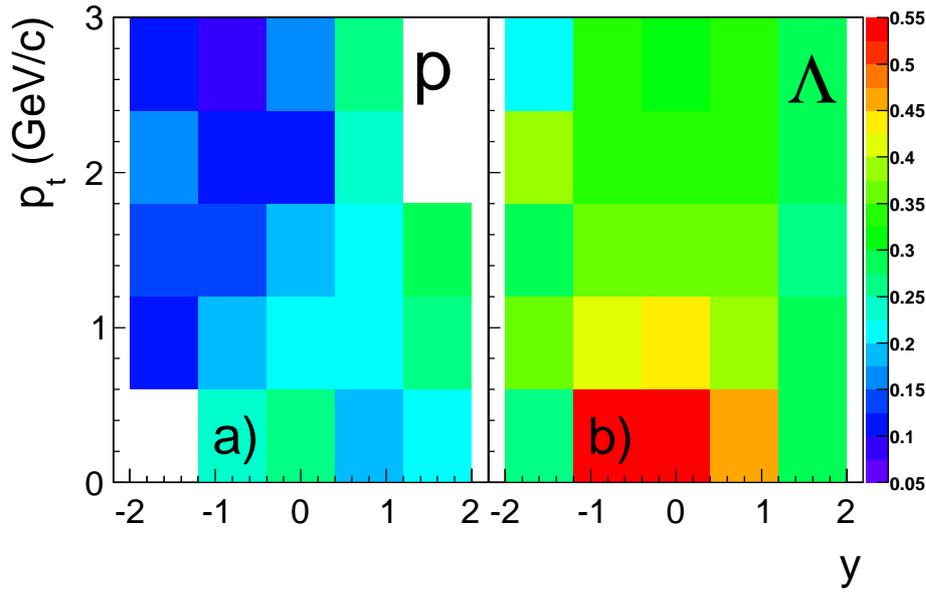}
\caption{\label{fig:feeddown} 
The fraction of protons ($F_{\rb{p}}(y,\pt)$) (a) and of \lam\
($F_{\Lambda}(y,\pt)$) (b) originating from feed-down as a function of
$y$ and \pt.}
\end{figure}
%

\section{Results} 

In total 70920 \lamp\ pair candidates, corresponding to 17520 real
\lamp\ pairs (i.e. after correcting for purities and feed-down), with
\mbox{$\qinv < 0.2$~GeV} were measured.  Dividing the signal
distribution by the event-mixing background and correcting with the
purities and feed-down contribution according to
\Eqs{eq:meascf}{eq:corrcf} yields the final \lamp\ correlation
function as shown in \Fi{fig:plamcf}.  The correlation function
exhibits a significant enhancement for small \qinv.  Such a
correlation would be expected as the effect of the strong interaction
between the proton and the \lam.

\subsection{Fit to theoretical calculation}

%
\begin{figure}[t]
\includegraphics[height=90mm]{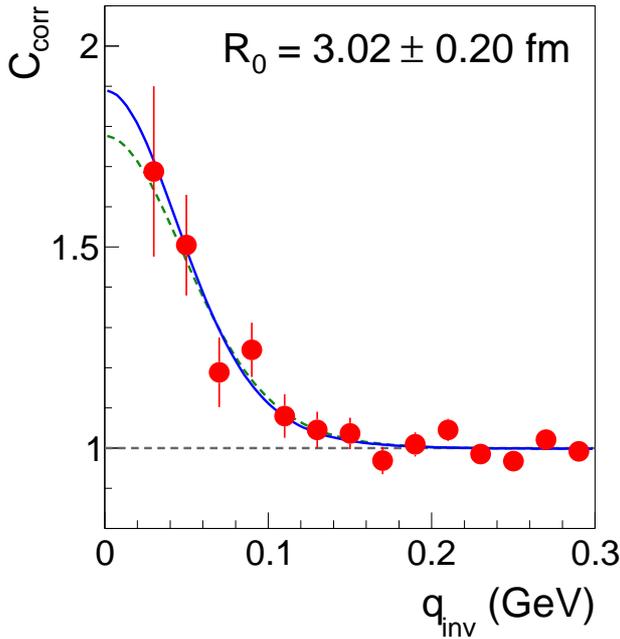}
\caption{\label{fig:plamcf} 
The corrected \lamp\ correlation function for central Pb+Pb reactions
at \sqrts~= 17.3~GeV, shown as a function of the invariant relative
momentum \qinv.  The data represent an average over the whole
acceptance of the NA49 experiment.  The lines display results of the
fit with a theoretical correlation function
\cite{LEDNICKY82,LEDNICKY05} (see text for details).  Only statistical
errors are shown.
}
\end{figure}
%

Since the shape of the momentum correlation function was shown by Wang
and Pratt to depend on the size of the emitting source \cite{WANG99},
it can be used to extract its radius.  The necessary prerequisite is a
quantitative knowledge of the \lamp\ interaction.  Here we use a
functional form of the theoretical correlation function $C_{\rb{th}}$
that is based on the model of Lednick\'{y} and Lyuboshitz
\cite{LEDNICKY82,LEDNICKY05}.  It employs an effective range
approximation of the S-wave \lamp\ interaction.  The source size is
required to be larger than the effective range of the interaction.
The strength of the interaction is defined by four parameters: the
effective ranges \dzeros\ (\dzerot) and the scattering lengths
\fzeros\ (\fzerot) for the singlet $S = 0$ (triplet $S = 1$) state.
In our fits we use values of \dzeros~=~2.92~fm, \dzerot~=~3.78~fm,
\fzeros~=~-2.88~fm, and \fzerot~=~-1.8~fm, as suggested in
\cite{WANG99}.  Under the assumption of unpolarized particle
production, the relative contribution of pairs in the singlet and the
triplet state is 1~:~3.  Furthermore, we use the same spherically
symmetric Gaussian spatial distribution $S(\vec{r})$, for both the
proton and the \lam\ source:
\begin{equation}
S(\vec{r}) = \exp \left(- \frac{x^{2} + y^{2} + z^{2}}
                               {2 \rgaus^{2}} \right) .
\end{equation}

%
\begin{figure}[t]
\includegraphics[height=120mm]{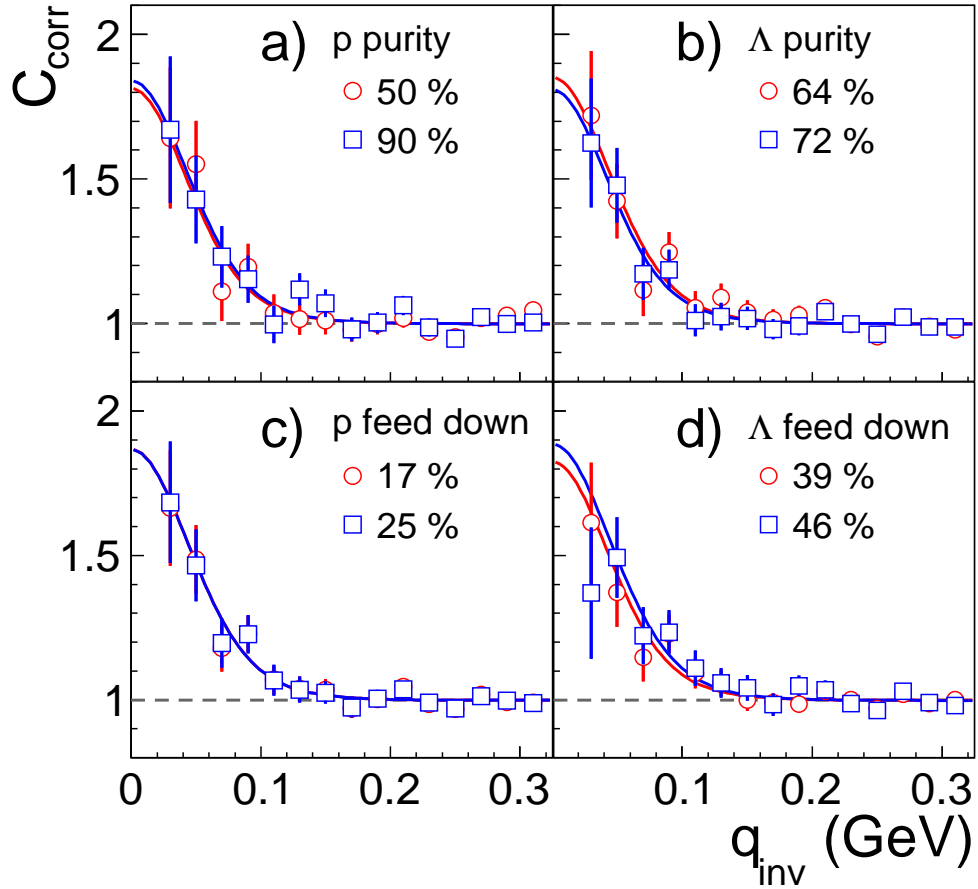}
\caption{\label{fig:systematics}
The corrected \lamp\ correlation function $C_{\rb{corr}}(\qinv)$ for
central Pb+Pb reactions at \sqrts~= 17.3~GeV shown as a function of
the invariant relative momentum \qinv\ for different purities of
protons (a) and \lam\ (b), as well as different feed-down
contributions to protons (c) and \lam\ (d) (see text).  The lines
display the results of fits with a theoretical correlation
function~\cite{LEDNICKY82,LEDNICKY05}.
}
\end{figure}
%

Then, the theoretical correlation function $C_{\rb{th}}$ assumes the
functional form as quoted in reference \cite{STARPLAM}.  By fitting it
to the data one obtains the effective radius parameter \rgaus\ and an
additional parameter $\lambda$ that takes possible reductions of the
height of the correlation into account:
\begin{equation}
\label{eq:fitcf}
C_{\rb{fit}}(\qinv) = \lambda \: (C_{\rb{th}}(\qinv) - 1) + 1 .
\end{equation}
These reductions occur, if the correction for the particle purities
and the feed-down, as defined in \Eq{eq:corrcf}, are insufficient.  If
both parameters are left to vary freely, the best fit is obtained for
$\rgaus = 2.70 \pm 0.60$(stat.)~fm and  $\lambda = 0.77 \pm
0.38$(stat.) (green dashed line in \Fi{fig:plamcf}).  The agreement
within errors of the fitted $\lambda$ value with unity underlines the
consistency of the correction procedure.  This justifies to fix
$\lambda = 1$ and to use only \rgaus\ as a free parameter, which
reduces the resulting error.  With this constraint we obtain
$\rgaus = 3.02 \pm 0.20$(stat.)~fm (blue solid line in
\Fi{fig:plamcf}).

\subsection{Systematic uncertainties}

Systematic errors arise both in the extraction of the correlation
function and in the theoretical model. The former uncertainties were
studied by making small changes in the analysis procedure.  By varying
the corresponding cuts on the measured energy loss used to identify
the primary protons, as well as the decay protons of the \lam\
candidates, the particle purities $P_{\rb{p}}(p_{\rb{p}})$ and
$P_{\lam}(y_{\lam}, p_{\rb{t},\lam})$ can be varied to a certain extent.
This changes the measured correlation function $C_{\rb{meas}}(\qinv)$.
However, after applying the appropriate correction factor $\langle
K(\qinv) \rangle$ the same corrected correlation function should be
obtained.  Figure~\ref{fig:systematics} shows comparisons of
$C_{\rb{corr}}(\qinv)$ for different proton (\Fi{fig:systematics},a))
and \lam\ purities (\Fi{fig:systematics},b)).  Even though the
correction factor changes quite dramatically ($\approx 45$~\% in case
of the proton purity), the resulting correlation functions agree quite
well.  The systematic error on the effective radius parameter \rgaus\
is derived by taking the maximal difference of \rgaus\ as obtained by
fits to the different correlation functions.  It is found that it
changes between $2.91 - 3.28$~fm ($2.98 - 3.29$~fm), if the proton
(\lam) purity is varied between $50 - 90$~\% ($64 - 72$~\%).
 
A similar study is performed by varying the feed-down contribution by
changing the cuts on the impact parameter $b_{\rb{x}}$ and
$b_{\rb{y}}$ ($\Delta x$ and $\Delta y$) of the proton tracks
(\Fi{fig:systematics},c) and \lam\ candidates
(\Fi{fig:systematics},d).  One finds that \rgaus\ changes maximally
between $3.02 - 3.13$~fm ($2.92 - 3.23$~fm), if the contributions to
the protons (\lam) from feed-down is varied between $17 - 25$~\% ($39
- 46$~\%).

The dependence of the radius parameter on the region in \qinv\ that is
used to determine the normalization constant $N$ (see \Eq{eq:meascf})
is investigated by varying the size and position of this region.  It
is found that \rgaus\ changes by maximally $\pm$0.11~fm.

By taking the quadratic sum of the different contributions a total
systematic error on \rgaus\ of $+0.44$~fm and $-0.16$~fm is
estimated.

%
%

Another source of systematic uncertainty of the extracted radius
parameter arises from the limited precision of the knowledge of
the scattering lengths and effective ranges used in the calculation of
the theoretical correlation function. Therefore, the fits were
repeated with \dzeros, \dzerot, \fzeros, and \fzerot\ taken from
\cite{COSY11,HIRES,BODMER88,SIBIRTSEV06,RIJKEN99,HAIDENBAUER05,LEDNICKY05}.  
The largest deviation from the \rgaus\ value obtained with the
standard parameter set \cite{WANG99} is observed when using the
parameters extracted by the COSY-11 collaboration \cite{COSY11}.
While in all other cases the difference is smaller than $\pm$0.1~fm,
it is found to be $+0.274$~fm for the COSY-11 parameters which is
still close to our statistical error.  Therefore we conclude that the
choice of the parameters describing the \lamp\ interaction has a
negligible effect on the final result.

\subsection{Energy dependence of the effective radius parameter}

%
\begin{figure}[t]
\includegraphics[height=100mm]{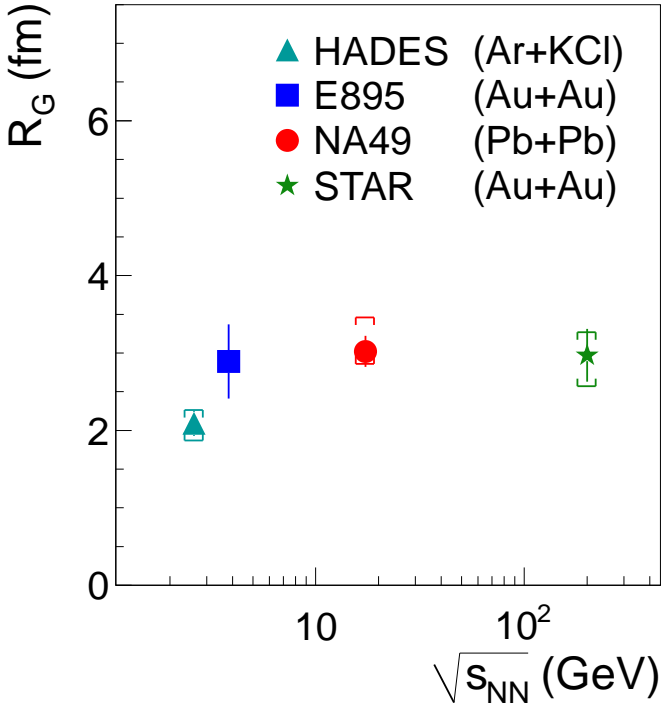}
\caption{\label{fig:rsqrts} 
The effective radius parameter \rgaus\ for \lamp\ correlations 
as a function of center of mass energy \sqrts.  The figure also
includes data on Ar+KCl collisions by the HADES collaboration
\cite{HADESPLAM} and on central Au+Au collisions by the E895
\cite{E895PLAM} and the STAR \cite{STARPLAM} experiments.  The
systematic errors are represented by the brackets.  Please note that
the STAR result corresponds to a slightly higher pair \mtavg\ than for
the NA49 measurement. Assuming the \mt~dependence shown
\Fi{fig:rmeanmt} the STAR data point would move up by $\approx 6$~\%,
if it was measured at the same \mtavg.
}
\end{figure}
%

The result of this study is compared to data at lower and higher
center-of-mass energies in \Fi{fig:rsqrts}.  Good agreement with the
effective radius parameters measured for central Au+Au collisions by
the E895 collaboration at \sqrts~= 3.83~GeV \cite{E895PLAM} and by the
STAR experiment at \sqrts~= 200~GeV \cite{STARPLAM} is observed,
indicating that there is no significant change from AGS to RHIC
energies.  A similar observation was made for correlations of
identical charged pions \cite{NA49HBT}.

%
\begin{figure}[t]
\includegraphics[height=100mm]{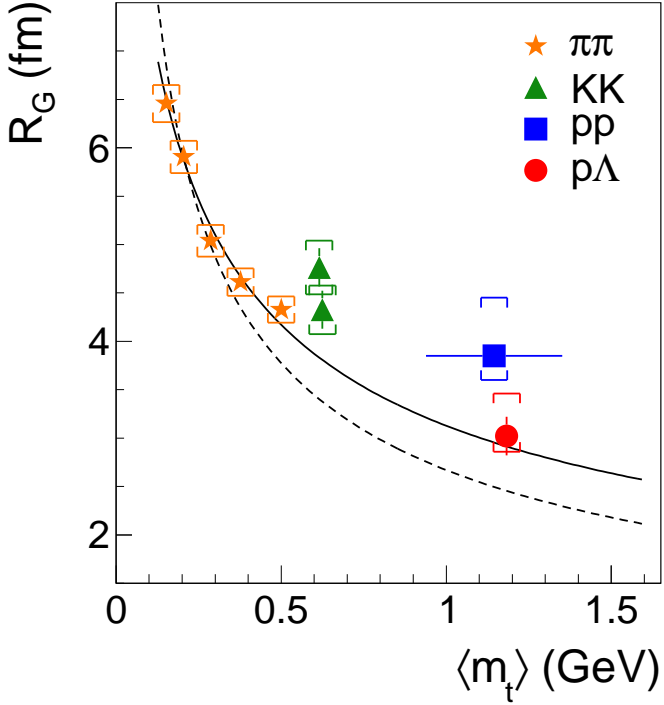}
\caption{\label{fig:rmeanmt}
The effective radius parameter \rgaus\ extracted from the correlation
functions of \pimin\pimin\ \cite{NA49HBT}, \kplus\kplus\ and
\kmin\kmin\ \cite{NA49KK}, pp \cite{NA49PP}, and \lamp\ pairs versus
\mtavg\ as measured by NA49 for central Pb+Pb collisions at \sqrts~=
17.3~GeV.  The systematic errors are represented by the brackets. For
explanation of curves see text.
}
\end{figure}
%

At even lower energies (\sqrts~= 2.61~GeV) the HADES collaboration
measured a significantly smaller effective radius parameter for \lamp\
correlations of \mbox{$\rgaus = 2.09 \pm 0.16$(stat.)~fm}
\cite{HADESPLAM}.  However, the fireball volume in the case of Ar+KCl
collisions is expected to be smaller than for Pb+Pb and Au+Au
collisions.  In fact, it was demonstrated in \cite{HADESPLAM}
that the measured \lamp\ radius parameter is dominated by the
reaction geometry and scales approximately as $A^{1/3}$, irrespective
of center-of-mass energy.  A comparable observation was previously
made for two-proton correlations in the target fragmentation
region \cite{WA80PP}.

\subsection{\mtavg\ dependence of the effective radius parameter}

Figure~\ref{fig:rmeanmt} shows a comparison of \rgaus, as determined
in this analysis from the \lamp\ correlation function, to radius
parameters derived from correlations of \pimin\ pairs \cite{NA49HBT}
and charged kaon pairs \cite{NA49KK}, as well as from two-proton
correlations \cite{NA49PP}, at different average transverse masses
\mtavg.  In case of the \pimin\pimin\  and KK correlations, \rgaus\
was calculated from the three-dimensional radius parameter components
as:
\begin{equation}
\label{eq:rzero}
\mbox{$\rgaus = (\rside \, \rout \, \rlong)^{1/3}$} .
\end{equation}

The decrease of the effective radius parameters with increasing
transverse mass is generally attributed to the presence of collective
flow in the fireball.  The measurement of the \lamp\ correlation
allows to extend this study to higher \mtavg.  In fact, \rgaus\ for
\lamp\ pairs is significantly smaller than the effective radius
parameter extracted for pions and kaons at lower \mtavg\ and is thus
in agreement with the expected behavior.  The dashed curve in
\Fi{fig:rmeanmt} corresponds to a simple $\propto \mtavg^{-1/2}$
dependence.  The solid line is based on the following \mt~dependences
of the three radius components, as suggested by hydrodynamical
approaches \cite{WIEDEMANN99,CHAPMAN95,MAKHLIN88}:
\begin{eqnarray}
\rside^{2} & = & \rgeo^{2} / (1 + (\mt/T) \: \etaf^{2}) 
                \label{eq:rside} \\
\rout^{2}  & = & \rside^{2} + \betat^{2} \: \Delta\tau^{2} \\
\rlong^{2} & = & \tauzero^{2} \: (T/\mt).
                \label{eq:rlong}
\end{eqnarray}
Here \rgeo\ is the transverse size of the particle source, $\betat =
v_{\rb{f}}/c$  the transverse flow velocity, \etaf\ the transverse
flow rapidity $\etaf = (1/2) \: \log (1 + v_{\rb{f}})/(1 -
v_{\rb{f}})$, $T$ the kinetic freeze-out temperature, \tauzero\ the
total lifetime of the source and $\Delta\tau$ the emission duration.
Under the assumption that all particle species freeze out from the
same expanding source, we calculate \rgaus\ from \Eq{eq:rzero}, using
\Eqr{eq:rside}{eq:rlong}.  The solid curve in \Fi{fig:rmeanmt}
corresponds to the parameter \etaf~=~0.8, \tauzero~=~0.6~fm/$c$,
$\Delta\tau$~=~3.4~fm/$c$ and $T$~=~90~MeV, as extracted by fits with
a blast wave model to the pion correlations \cite{NA49HBT}.  The
overall normalization has been adjusted to fit the data.  A reasonable
description of the effective radius parameters for most particle
species can thus be achieved, with the notable exception of the
two-proton correlation.


\section{Summary}

We report on the measurement of the \lamp\ correlation
function in momentum space for central Pb+Pb collisions at \sqrts~=
17.3~GeV.  The \lamp\ pairs exhibit a clear positive correlation for
small relative momenta.  By comparison to a calculated correlation
function a one-dimensional Gaussian source size parameter of
\mbox{$\rgaus = 3.02 \pm 0.20$(stat.)$^{+0.44}_{-0.16}$(syst.)~fm} is
determined.  This value is in good agreement with measurements for
Au+Au collisions at lower and higher center-of-mass energies.  The
\mtavg~dependence of the effective \lamp\ radius parameter follows the
expectation for an expanding source as described by hydrodynamics,
when compared to other two-particle correlation results.


\section*{Acknowledgments}

This work was supported by
the US Department of Energy Grant DE-FG03-97ER41020/A000,
the Bundesministerium fur Bildung und Forschung, Germany (06F~137),
the German Research Foundation (grant GA 1480/2-1),
the Polish Ministry of Science and Higher Education (1~P03B~006~30,
1~P03B~127~30, 0297/B/H03/2007/33, N~N202~078735, N~N202~078738,
N~N202~204638),
the Hungarian Scientific Research Foundation (T068506), 
the Bulgarian National Science Fund (Ph-09/05),
the Croatian Ministry of Science, Education and Sport (Project
098-0982887-2878) and
Stichting FOM, the Netherlands.



\end{document}